\documentclass[12pt]{article}
\usepackage[english,russian]{babel}
\usepackage{latexsym}
\usepackage{amsthm}
\usepackage{graphicx}
\usepackage{subeqnarray}
\usepackage{amsmath}
\usepackage{amsfonts}
\usepackage{amscd}
\usepackage{amsthm}
\usepackage{amssymb}
\usepackage{latexsym}
\newcommand{\bit}{\begin{itemize}}
\newcommand{\eit}{\end{itemize}}
\newcommand{\rd}{{\rm{d}}}
\def\ri{{\rm i}}

\begin{document}

\begin{center}
{\large\bf Damaskinsky E.V.$^{a}$,\, Sokolov M.A.$^{b}$}
\bigskip

{\Large\bf On differential operators for bivariate Chebyshev polynomials
\footnote{The work is supported by RFBR under the grant 15-01-03148}}
\end{center}

\vspace{0.2cm}

$^a$ Math. Dept. Military Engineering Institute. VI(IT). Saint Petersburg and Petersburg Department
of the Steklov Mathematical Institute,  evd@pdmi.ras.ru.
\bigskip

$^b$ Peter the Great St. Petersburg Polytechnic University and Military Telecommunications
Academy (MTA), masokolov@gmail.com.
\vspace{0.5cm}

\begin{quote} {\bf Abstract} We construct the differential operators for which bivariate Chebyshev
polynomials of the first kind, associated with simple Lie algebras $C_2$ and $G_2$, are eigenfunctions.
\end{quote}
\vspace{0.5cm}

{\large\bf 1.}
In these notes, we obtain differential operators for which bivariate Chebyshev polynomials of the
first kind, associated with the root systems of the simple Lie algebras $C_2$ and $G_2$, are
eigenfunctions. For the case of bivariate Chebyshev polynomials,
associated with the Lie algebra $A_2$, such operators were obtained in the well known
Koornwinder's work \cite{K1}.

Chebyshev polynomials in several variables are natural generalizations of the classical Cheby\-shev
polynomials in one variable (see, for example \cite{Ri}). The polynomials of the first kind can be
defined in the following manner.

Denote by $R$ a reducible system of roots for a simple Lie algebra $L$.
A system of roots is a set of vectors in $d$-dimensional Euclidean space $E^d$
with a scalar product $(.,.)$. This system is completely determined by a basis of simple roots
$\alpha_i,\,i=1,..,d$ and by a group of reflections of $R$ called a Weyl group $W(R)$.
Generating elements of the Weyl group $w_i,\,i=1,..,d$ acts on any vector $x\in E^d$  according to the formula
\begin{equation}
\label{0-1}
w_i\,x=x-\frac{2(x,\alpha_i)}{(\alpha_i,\alpha_i)}\alpha_i.
\end{equation}
In particular, if $x = \alpha_i$ we obtain from (\ref{0-1}) $w_i\,\alpha_i =-\alpha_i$.
A system of roots $R$ is closed under the action of related Weyl group $W(R)$.

To any root $\alpha$ from the system $R$ corresponds the coroot
$$
\alpha^{\vee}=\frac{2\alpha}{(\alpha,\alpha)}.
$$
For the basis of the simple coroots $\alpha^{\vee}_i,\,i=1,..,d$ one can define the dual
basis of fundamental weights
$\lambda_i,\,i=1,..,d$
$$
(\lambda_i,\alpha^{\vee}_j)=\delta_{ij}
$$
(we identify the dual space ${E^d}^*$ with $E^d$).
The bases of roots and weights are related by the linear transformation
\begin{equation}
\label{0-2}
\alpha_i=\sum_j C_{ij}\lambda_j, \quad C_{ij}=\frac{2(\alpha_i,\alpha_j)}{(\alpha_j,\alpha_j)},
\end{equation}
where $C$ is the Cartan matrix of the Lie algebra $L$.

For any Lie algebra $L$ with related sistem of roots $R$  and Weyl group $W(R)$,
an orbit function $\Phi_{\bf n}({\boldsymbol{\phi}})$
is defined as
\begin{equation}
\label{0-3}
T_{\bf n}^{L}({\boldsymbol{\phi}}) = \frac1{|W(R)|}
\sum\limits_{w\in {\mbox{\footnotesize W(R)}}}e^{\rm i(\emph{w}\,{\bf n},{\boldsymbol{\phi}})}.
\end{equation}
In the formula (\ref{0-3}) $|W(R)|$ is a number of elements in a group $W(R)$,
$\bf n$ is expressed in the basis of fundamental weights $\{\lambda_i\}$ and $
{\boldsymbol{\phi}}$ is expressed in the dual basis of coroots $\{\alpha^\vee_i\}$
$$
{\bf n}=\sum_{i=1}^d\,n_i\lambda_i \quad n_i\in Z, \quad  {\boldsymbol{\phi}}=
\sum_{i=1}^d\,\phi_i\alpha_i^\vee \quad \phi_i\in [0,2\pi).
$$
Obviously $T_{\bf n}^{L}({\boldsymbol{\phi}})$ is a $W(R)$-invariant function because of
$$
T_{\tilde w\,\bf n}^{L}({\boldsymbol{\phi}})=T_{\bf n}^{L}({\boldsymbol{\phi}}), \,\,\forall \tilde w\,\in W(R).
$$
Then we define the  new variables  $x_i$ (generalized cosines) by the relations
\begin{equation}
\label{0-4}
x_i\,=\,T_{{\bf e}_i}({\boldsymbol{\phi}}),\quad {\bf e}_i
= (\overbrace{0,..,0}^{i-1},1,\overbrace{0,..,0}^{d-i}).
\end{equation}
It is shown in the works  \cite{K1,K2,H,HW,B,KP} that the function
$T_{\bf n}({\boldsymbol{\phi}})$
defined by the formula (\ref{0-3}) with non-negative integer $n_i$ from
${\bf n} = (n_1,...,n_d)$ can be expressed in the terms of $x_i$. This function
gives us up to a normalization the multivariate Chebyshev polynomials $T_{n_1,...,n_d}$ of the first kind.
\bigskip

{\large\bf 2.}
The simplest example of the above construction is the classical Chebyshev polynomials associated
with the Lie algebra $A_1$. The related Weyl group consists from the identical element $w_0$ and
the reflection of the single positive root $w_1\lambda =-\lambda$. In this case the definition
(\ref{0-3}) gives

\begin{equation}
\label{0-5}
T_n({\phi})=\frac1{2}(e^{\rm i n\phi}+e^{-\rm i n\phi}) = \cos{n\phi},\quad x = T_1({\phi}) = \cos{\phi}.
\end{equation}

To derive the differential operator(s) for which the classical polynomials of the first
kind $T_n(x)$ are eigenfunction we firstly write out the differential equation for $\cos{n\phi}$
\begin{equation}
\label{0-6}
\frac{\rd^2\cos{n\phi}}{\rd\phi^2} + n^2\cos{n\phi} = 0.
\end{equation}
It follows from (\ref{0-6}) that desired operator in terms of the  angle variable $\phi$ has the form

\begin{equation}
\label{0-7}
L^{(A_1)}(\phi)=\frac{\rd^2}{\rd\phi^2}.
\end{equation}
Changing the variable $\cos{\phi} \rightarrow x$ in (\ref{0-7}) we obtain the well known operator in terms of $x$
\begin{equation}
\label{0-8}
L^{(A_1)}(x) = (1-x^2)\frac{\rd^2}{\rd x^2}-x\frac{\rd }{\rd x}.
\end{equation}
\bigskip

{\large\bf 3.}
Now we turn to the generalized cosine associated with the Lie algebra $A_2$.
At the first step we find the orbit function related to the algebra $A_2$.
The root system of this algebra has two fundamental roots $\alpha_1,\,\alpha_2$ and includes the
positive root $\alpha_1 + \alpha_2$ together with their reflections. The action of generating
elements $w_1,w_2$ of the Weyl group $W(A_2)$ on the fundamental roots are given by the formulas
$$w_1\alpha_1=-\alpha_1,\quad w_1\alpha_2=\alpha_1+\alpha_2,\quad w_2\alpha_1=\alpha_1+\alpha_2,
\quad w_2\alpha_2=-\alpha_2.$$
Taking into account (\ref{0-2}) and explicit form of the Cartan matrix $C(A_2)$ (see, for example \cite{Hu})
we obtain the action of $w_1,w_2$ on the fundamental weights
\begin{equation}
\label{0-9}
w_1\lambda_1=\lambda_2-\lambda_1,\quad w_1\lambda_2=\lambda_2,\quad w_2\lambda_1=
\lambda_1,\quad w_2\lambda_2=\lambda_1-\lambda_2.
\end{equation}
The action of the other group elements on the fundamental weights is determined by their representation in terms of the generating elements
\begin{equation}
\label{0-10}
w_3=w_1w_2,\quad w_4=w_2w_1,\quad w_5=w_1w_2w_1,\quad w_0=e.
\end{equation}
Using these formulas, the definition (\ref{0-3}) and the notation
$${\bf n} = m\lambda_1 + n\lambda_2,\quad {\boldsymbol{\phi}}=\phi\alpha^{\vee}_1+\psi\alpha^{\vee}_2$$
we find the $W(A_2)$-invariant function of two variables
\begin{multline}\label{0-11}
T_{m,n}(\phi,\psi)=\\
e^{{\rm i}m\phi}e^{{\rm i}n\psi}+e^{{\rm i}m(\psi-\phi)}e^{{\rm i}n\psi}+e^{{\rm i}m\phi}e^{{\rm i}n(\phi-\psi)}
+e^{{\rm i}m(\psi-\phi)}e^{-{\rm i}n\phi}+e^{-{\rm i}m\psi}e^{{\rm i}n(\phi-\psi)}+
e^{-{\rm i}m\psi}e^{-{\rm i}n\phi}.
\end{multline}
The normalization factor was omitted in (\ref{0-11})  because it is not essential for our purpose.

At the second step we find differential operators for which the orbit functions $T_{m,n}(\phi,\psi)$
for any $m,n$ are the eigenfunctions
$$
L_N(T_{m,n})=E_{m,n}T_{m,n}.
$$
The form of the orbit function implies that the action of the operator $L_N$ on each exponent
from  (\ref{0-11}) must gives us the same eigenvalues $E_{m,n}$ for any $m,\,n$. For this reason we
search the operators of the form
\begin{equation}
\label{0-12}
L_N^{(A_2)}(\phi ,\psi)=\sum_{k=0}^{N}a_{k}\frac{\partial^N}{\partial\phi^{(N-k)}\partial\psi^k},
\end{equation}
with real constant coefficients  $a_{k},\, k = 0,..,N$.

Let us act by the operator $L_N^{(A_2)}(\phi ,\psi)$ on $T_{m,n}$ and write out the chain of
equalities of coefficients at the each exponent of (\ref{0-11})
$$
\sum_{k=0}^{N}a_{k}m^{N-k}n^k =  \sum_{k=0}^{N}a_{k}(-m)^{N-k}(m+n)^k =
\sum_{k=0}^{N}a_{k}(m+n)^{N-k}(-n)^k =
$$
$$
\sum_{k=0}^{N}a_{k}(-m-n)^{N-k}(m)^k = \sum_{k=0}^{N}a_{k}n^{N-k}(-m-n)^k =
\sum_{k=0}^{N}a_{k}(-n)^{N-k}(-m)^k.
$$
Some conclusions about the properties of the coefficients $a_{k}$ can be made directly from the
form of the sums. For example, changing the summation index in the last sum of the chain
$k\rightarrow N-k$ and compare this sum with the first one we conclude that $a_{k}=a_{N-k}$
for the even $N$, and $a_{k}=-a_{N-k}$ for the odd $N$.

To calculate the coefficients $a_{k}$ in the explicit form it is necessary to solve some equation
systems which arise from equalization of coefficients at the same monomials $m^pn^q$ in the above
chain. It is convenient to reformulate this problem as a problem of calculation of the vector
$$
V_{N+1} = (a_{0},a_{1},...,a_{N})
$$
which is a common eigenvector with the eigenvalue $1$ of the matrices related to the equation
systems under consideration.

Consider for example the first equality from the chain. We can write the following equation
\begin{equation}
\label{0-13}
{\tt M}_1V_{N+1}= E_{N+1}V_{N+1} = V_{N+1},
\end{equation}
where $E_{N+1}$ is the unit $(N+1)\times(N+1)$ matrix, $M_1$ is the lower triangular matrix of the
same degree with the nonzero matrix elements
\begin{equation}
\label{0-14}
({{\tt M}_1})_{ij}= (-1)^{j+1}{{N+1-j}\choose {N+1-i}},\quad i,j = 1,..N+1,
\end{equation}
where ${{j}\choose {i}}$ is the binomial coefficient.
The equality of the first and third sums gives us the equation
\begin{equation}
\label{0-15}
{\tt M}_2V_{N+1}=V_{N+1},
\end{equation}
where $M_2$ is the upper  triangular matrix with the nonzero matrix elements
$$
({{\tt M}_2})_{ij}= (-1)^{j+1}{{j-1}\choose {i-1}},\quad i,j = 1,..N+1.
$$
By the same manner we obtain the matrices $M_i,\,\,i=3,4,5,$ from the above equalities.
It can be easily checked that these matrices are connected ${\tt M}_1,{\tt M}_2$ by the following formulas
$$ {\tt M}_3={\tt M}_1{\tt M}_2,\quad {\tt M}_4={\tt M}_2{\tt M}_1,\quad {\tt M}_5=
{\tt M}_1{\tt M}_2{\tt M}_1,\quad {\tt M}_0=E_{N+1}.$$
Moreover, under the correspondence $w_i\sim {\tt M}_i$ we reproduce the multiplication table of
the Weyl group $W(A_2)$ including the equalities
$$
{\tt M}_1^2={\tt M}_2^2={\tt M}_5^2={\tt M}_3^3={\tt M}_4^3=E_{N+1},\quad {\tt M}_3^2={\tt M}_4,\,{\tt M}_4^2=
{\tt M}_3.
$$
It follows from the above that the homomorphism $w_i\rightarrow {\tt M}_i,\,i=0,..,5$,
${\tt M}_0=E_{N+1}=w_0$ realizes faithful $(N+1)$-dimensional representation of the Weyl group $W(A_2)$.
Since the matrices ${\tt M}_1$ and ${\tt M}_2$ are the images of the generators for the Weyl group $W(A_2)$,
we can calculate the joint eigenvectors only for these two matrices.

Joint solution of (\ref{0-13}) and (\ref{0-15}) in the cases $N=2,3$ gives us the following
result
\begin{equation}
\label{0-16}
N=2,\quad V^{A_2}_{3} = (1,1,1),\quad N=3,\quad V_{4}^{A_2} = (2,3,-3,-2).
\end{equation}
The related independent operators in the angle variables with their spectrums have the forms
\begin{equation}
\label{0-17}
L^{A_2}_{3}=\partial^2_{\phi^2}+\partial^2_{\phi\psi}+
\partial^2_{\psi^2},\quad E^{A_2}_{3}(m,n)=m^2+mn+n^2,
\end{equation}
\begin{equation}
\label{0-18}
L^{A_2}_{4}=2\partial^3_{\phi^3}+3\partial^3_{\phi^2\psi}-3\partial^3_{\phi\psi^2}-
2\partial^3_{\psi^3},
\quad E^{A_2}_{4}(m,n)=2m^3+3m^2n-3mn^2-2n^2.
\end{equation}
High degree operators can be constructed as
$$
L = P(L^{A_2}_{3},L^{A_2}_{4})
$$
where $P$ is any polynomial in two variables.
\bigskip

{\large\bf 4.}
At the last step it is necessary to replace the angle variables $(\phi,\psi)$ by $(x,y)$ which
are defined according to the relation (\ref{0-4}) as
\begin{equation}
\label{2-11}
x = \frac{1}{2}T_{1,0} = e^{i\phi}+e^{i(\psi-\phi)}+e^{-i\psi},
\end{equation}
\begin{equation}
\label{2-12}
y = \frac{1}{2}T_{0,1} = e^{i\psi}+e^{i(\phi-\psi)}+e^{-i\phi}.
\end{equation}
This routine procedure in the case $N=2$ gives us the operator
\begin{equation}
\label{2-13}
L_{3}^{A_2}=(x^2-3y)\frac{\partial^2}{\partial x^2}+(xy-9)\frac{\partial^2}{\partial x\partial y}+(y^2-3x)\frac{\partial^2}{\partial y^2}+x\frac{\partial}{\partial x}+y\frac{\partial}{\partial y}.
\end{equation}
The bivariate Chebyshev polynomials of the first kind associated with the Lie algebra $A_2$ are
eigenvectors of $L_{3}^{A_2}$ with eigenvalues defined by (\ref{0-17}).
The operator (\ref{2-13}) was obtained for the first time by T.  Koornwinder in the well known
work \cite{K1}. Our calculation method, presented above, is different from the method used in \cite{K1}.
\bigskip

{\large\bf 5.}
Here we use the same calculation scheme as above for the case of the polynomials,
associated with the Lie algebra $C_2$. The root system of the algebra $C_2$ has two fundamental
roots $\alpha_1,\,\alpha_2$ and includes the positive root $\alpha _1+\alpha_2,\,2\alpha _1+\alpha_2$
and their reflections. The action of generating elements $w_1,w_2$ of the Weyl group $W(A_2)$ on
the fundamental roots are given by the formulas
$$
w_1\alpha_1=-\alpha_1,\quad w_1\alpha_2=2\alpha_1+\alpha_2,\quad w_2\alpha_1=\alpha_1+\alpha_2,
\quad w_2\alpha_2=-\alpha_2,
$$
$$
w_1\lambda_1=\lambda_2-\lambda_1,\quad w_1\lambda_2=\lambda_2,\quad w_2\lambda_1=\lambda_1,
\quad w_2\lambda_2=2\lambda_1-\lambda_2.
$$
The action of the other group elements on the fundamental weights is determined by their
representation in terms of the generating elements
\begin{equation}\label{cc-1}
w_3=w_1w_2,\quad w_4=w_2w_1,\quad w_5=w_1w_2w_1,\quad w_6=w_2w_1w_2,\quad w_7=(w_1w_2)^2,\quad e=w_0.
\end{equation}
Using the above formulas we  obtain the following $W(C_2)$-invariant orbit function
\begin{equation}\label{cc-3}
\begin{split}
T_{m,n}^{C_2}(\phi,\psi) &= e^{2\pi\ri(m\phi+n\psi)}+e^{2\pi\ri(m(\psi-\phi)+n\psi))}+
e^{2\pi\ri(m\phi+n(2\phi-\psi))}+e^{2\pi\ri(m(\psi-\phi)+n(-2\phi+\psi))}+  {}\\
&+ e^{2\pi\ri(m(\phi-\psi)+n(2\phi-\psi))}+e^{2\pi\ri(-m\phi+n(-2\phi+\psi))}+
e^{2\pi\ri(m(\phi-\psi)-n\psi)}+e^{2\pi\ri(-m\phi-n\psi)}.
\end{split}
\end{equation}
The action of the operator (\ref{0-12}) on $T_{m,n}^{C_2}(\phi,\psi)$ produces coefficients at
each exponent of (\ref{cc-3}). The condition of equality of these  coefficients gives us the
following independent relations
$$
\sum_{k=0}^{N}a_{k}m^{N-k}n^k =  \sum_{k=0}^{N}a_{k}(m)^{N-k}(-m-n)^k =
\sum_{k=0}^{N}a_{k}(m+2n)^{N-k}(-n)^k =
$$
$$
\sum_{k=0}^{N}a_{k}(m+2n)^{N-k}(-m-n)^k = \sum_{k=0}^{N}a_{k}(-m)^{N-k}(-n)^k.
$$
It follows from the equality of the first and last sums that the coefficients $a_{k}$ are nonzero
only for the even $N$. In this case the matrix elements of the matrices
${\tt M}_i,\,\,i=1,2$ have the form
$$
({{\tt M}_1})_{ij}= (-1)^{j+1}{{N+1-j}\choose {N+1-i}},\quad
({{\tt M}_2})_{ij}= (-1)^{j+1}2^{j-i}{{j-1}\choose {i-1}},\quad i,j = 1,..N+1.
$$
These matrices are commutative
$$
[{\tt M}_1,{\tt M}_2]=0,\quad{\tt M}_1^2={\tt M}_2^2=E_{N+1}.
$$
Besides $M_i,\,\,i=1,2$ there is only one independent matrix $M_3$
$$
{\tt M}_3 ={\tt M}_1{\tt M}_2.
$$
Coordinates $a_{k}$ of any joint eigenvectors with unit eigenvalues of the matrices
${\tt M}_i,\,\,i=1,2$ give us the coefficients of the operator $L_N^{(C_2)}$ from (\ref{0-12}).
For the cases $N=2,4$ we obtain the following result
\begin{equation}\label{0-19}
N=2,\quad V^{C_2}_{3} = (1,2,2),\quad N=4,\quad V_{5a}^{C_2} = (1,4,1,0,0),\quad V_{5b}^{C_2} = (0,0,1,2,1).
\end{equation}
The related independent operators in the angle variables with their spectrums have the forms
\begin{equation}\label{0-20}
L^{C_2}_{3}=\partial^2_{\phi^2}+2\partial^2_{\phi\psi}+2\partial^2_{\psi^2},
\qquad E^{C_2}_{3}(m,n)=m^2+2mn+2n^2,
\end{equation}
\begin{equation}\label{0-21}
L^{C_2}_{5a}=\partial^4_{\phi^4}+4\partial^4_{\phi^3\psi}+\partial^4_{\phi^2\psi^2},
\qquad E^{C_2}_{5a}(m,n)=m^2(m^2+4mn+n^2).
\end{equation}
\begin{equation}\label{0-22}
L^{C_2}_{5b}=\partial^4_{\phi^2\psi^2}+2\partial^4_{\phi\psi^3}+\partial^4_{\psi^4},
\qquad E^{C_2}_{5b}(m,n)=n^2(m+n)^2.
\end{equation}
\bigskip

{\large\bf 6.}
Transition from the angle coordinates to Descartes ones are given by the relations
(see, for example, \cite{DKS})
\begin{eqnarray}
\label{cc-6}
x&=&\frac{1}{2}T_{1,0}^{C_2}=e^{2\pi\ri\phi}+e^{-2\pi\ri\phi}+ e^{2\pi\ri(\phi-\psi)}+
e^{-2\pi\ri(\phi-\psi)}, \label{cc-5}\\
y&=&\frac{1}{2}T_{0,1}^{C_2}=e^{2\pi\ri\psi}+e^{-2\pi\ri\psi}+e^{2\pi\ri(2\phi-\psi)}+
e^{-2\pi\ri(2\phi-\psi)}.
\end{eqnarray}
For the case (\ref{0-20}) we obtain
\begin{equation}
\label{0-23}
L^{C_2}(x,y) = (x^2-2y-8)\frac{\partial^2}{\partial x^2}+2x(y-4)\frac{\partial^2}{\partial x\partial y}+2(y^2+4y-2x^2)\frac{\partial^2}{\partial y^2}+x\frac{\partial}{\partial x}+2y\frac{\partial}{\partial y}.
\end{equation}
\bigskip

{\large\bf 7.}
To finish these brief notes we consider the case of the polynomials, associated with the Lie algebra $G_2$.
The root system of the algebra $G_2$ has two fundamental roots $\alpha_1,\,\alpha_2$ and includes the
positive roots $\alpha _1+\alpha_2,\,2\alpha _1+\alpha_2,\,3\alpha _1+\alpha_2,\,3\alpha _1+2\alpha_2$
and their reflections. The action of generating elements $w_1,w_2$ of the Weyl group $W(A_2)$ on
the fundamental roots are given by the formulas
$$
w_1\alpha_1=-\alpha_1,\quad w_1\alpha_2=3\alpha_1+\alpha_2,\quad w_2\alpha_1=\alpha_1+\alpha_2,
\quad w_2\alpha_2=-\alpha_2,
$$
$$
w_1\lambda_1=\lambda_2-\lambda_1,\quad w_1\lambda_2=\lambda_2,\quad w_2\lambda_1=\lambda_1,
\quad w_2\lambda_2=2\lambda_1-\lambda_2.
$$
The action of the other group elements on the fundamental weights is determined by their representation
in terms of the generating elements
$$
w_3=w_1w_2,\, w_4=w_2w_1,\, w_5=w_2w_1w_2,\, w_6=w_1w_2w_1,\, w_7=(w_1w_2)^2,
$$
$$
\!w_8\!=\!(w_2w_1)^2,\, w_9\!=\!w_2(w_1w_2)^2,\, w_{10}\!=\!w_1(w_2w_1)^2,\,
w_{11}\!=\!(w_1w_2)^3,\, w_{0}\!=\!e.$$
Using these formulas and definition (\ref{0-3}) we obtain the following $W(G_2)$-invariant orbit function
\begin{multline}
\label{4-01}
T^{G_2}_{m,n} = e^{2\pi\ri(m\phi+n\psi)}+ e^{2\pi\ri(m(-\phi +\psi)+ n(-3\phi +2\psi))}+
e^{2\pi\ri(m(2\phi -\psi) + n(3\phi -\psi))}+ \\
 e^{2\pi\ri(-(m\phi + n\psi))}+ e^{2\pi\ri(-(m(-\phi +\psi)+n(-3\phi +2\psi)))} +
 e^{2\pi\ri(-(m(2\phi -\psi)+n(3\phi -\psi)))} + \\ e^{2\pi\ri(m\phi +n(3\phi-\psi))}+
 e^{2\pi\ri(m(-\phi + \psi)+n\psi)}+ e^{2\pi\ri(m(2\phi-\psi) + n(3\phi -2\psi))}+\\
 e^{2\pi\ri(-(m\phi  +n(3\phi-\psi)))} +
  e^{2\pi\ri(-(m(-\phi + \psi)+n\psi))} + e^{2\pi\ri(-(m(2\phi-\psi) + n(3\phi -2\psi)))}.
\end{multline}
The action of the operator (\ref{0-12}) on $T_{m,n}^{G_2}(\phi,\psi)$ produces coefficients at each
exponent of (\ref{4-01}). The condition of equality of these  coefficients gives us the following
independent relations
$$
\sum_{k=0}^{N}a_{k}m^{N-k}n^k = \sum_{k=0}^{N}a_{k}(m)^{N-k}(-m-n)^k = \sum_{k=0}^{N}a_{k}(m+3n)^{N-k}(-n)^k =
$$
$$
\sum_{k=0}^{N}a_{k}(2m+3n)^{N-k}(-m-n)^k = \sum_{k=0}^{N}a_{k}(m+3n)^{N-k}(-m-2n)^k
=\sum_{k=0}^{N}a_{k}(2m+3n)^{N-k}(-m-2n)^k.
$$
Equality of the first and the second sums gives us the matrix ${\tt M}_1$ which is the same as in the
$A_2$ and $C_2$ cases (\ref{0-14}). Equality of the first and the second sums gives us the matrix
${\tt M}_2$
$$
({{\tt M}_1})_{ij}= (-1)^{j+1}{{N+1-j}\choose {N+1-i}},\quad
({{\tt M}_2})_{ij}= (-1)^{j+1}3^{j-i}{{j-1}\choose {i-1}},\quad i,j = 1,..N+1.
$$
The remaining matrices are
$$
{{\tt M}_3}={{\tt M}_1}{{\tt M}_2},\quad {{\tt M}_4}={{\tt M}_2}{{\tt M}_1},\quad {{\tt M}_5}=
{{\tt M}_1}{{\tt M}_2}{{\tt M}_1}={{\tt M}_2}{{\tt M}_1}{{\tt M}_2}.
$$
Coordinates $a_{k}$ of any joint eigenvectors with unit eigenvalues of the matrices
${\tt M}_i,\,\,i=1,2$ give us the coefficients of the operator $L_N^{(G_2)}$ from (\ref{0-12}).
For the cases $N=2$ we obtain the following result (there are no solutions for the odd cases)
\begin{equation}\label{g-1}
N=2,\quad V^{G_2}_{3} = (1,3,3).
\end{equation}
The related independent operator in the angle variables with its spectrum has the form
\begin{equation}\label{g-2}
L^{G_2}_{3}=\partial^2_{\phi^2}+3\partial^2_{\phi\psi}+3\partial^2_{\psi^2},
\qquad E^{G_2}_{3}(m,n)=m^2+3mn+3n^2,
\end{equation}
Calculations in the cases $N=4,6$ give us only
$L^{G_2}_{5}=\left(L^{G_2}_{3}\right)^2,\,L^{G_2}_{7}=\left(L^{G_2}_{3}\right)^3$.
\bigskip

{\large\bf 8.}
Transition from the angle coordinates to Descartes ones is given by the relations
\begin{eqnarray}\label{cc-6}
x&=&\frac{1}{2}T_{1,0}^{G_2}=e^{2\pi\ri(\phi)}+
e^{2\pi\ri(-\phi +\psi)}+
e^{2\pi\ri(2\phi -\psi)}+
e^{2\pi\ri(-\phi)}+
e^{2\pi\ri(-(-\phi +\psi))} +
e^{2\pi\ri(-(2\phi -\psi))},\\
y&=&\frac{1}{2}T_{0,1}^{G_2}= e^{2\pi\ri(\psi)}+
 e^{2\pi\ri(-3\phi +2\psi)}+
 e^{2\pi\ri(3\phi -\psi)}+
 e^{2\pi\ri(-\psi)}  +
 e^{2\pi\ri(\phi -2\psi)} +
 e^{2\pi\ri(-3\phi +\psi)} .
\end{eqnarray}
For the case (\ref{g-2}) we obtain
$$
L^{G_2}(x,y) = (x^2-3x-y-12)\frac{\partial^2}{\partial x^2}+(3xy-6x^2+12y+36)\frac{\partial^2}{\partial x\partial y}+$$
$$+(3y^2+9y-3x^3+9xy+27x)\frac{\partial^2}{\partial y^2}+x\frac{\partial}{\partial x}+3y\frac{\partial}{\partial y}.$$

\end{document}